\documentclass[journal,12pt,onecolumn,draftclsnofoot,]{IEEEtran}

\usepackage[cmex10]{amsmath}
\usepackage[utf8]{inputenc}
\usepackage{multirow}
\usepackage{soul}
\usepackage{algorithm}
\usepackage{algpseudocode}
\usepackage{graphicx}
\usepackage{dsfont}
\usepackage{xcolor}
\usepackage{cite}

\usepackage[utf8]{inputenc}
\usepackage{mathtools}
\usepackage[mathscr]{euscript}
\usepackage{amsmath}
\usepackage{amssymb}
\usepackage{amsfonts}
\usepackage{amssymb}
\usepackage{amsmath}
\usepackage{verbatim}
\usepackage[T1]{fontenc}
\usepackage[utf8]{inputenc}
\usepackage{latexsym}
\usepackage{enumerate}
\usepackage{indentfirst}
\usepackage{calligra}
\usepackage{ulem}
\usepackage{graphicx}
\usepackage{array}
\usepackage{float}
\usepackage{bbm}

\usepackage{geometry}

\usepackage{mleftright}
\usepackage{colortbl}
\makeatletter
\let\old@ps@headings\ps@headings
\let\old@ps@IEEEtitlepagestyle\ps@IEEEtitlepagestyle
\def\psccfooter#1{%
 \def\ps@headings{%
 \old@ps@headings%
 \def\@oddfoot{\strut\hfill#1\hfill\strut}%
 \def\@evenfoot{\strut\hfill#1\hfill\strut}%
 }%
 \def\ps@IEEEtitlepagestyle{%
 \old@ps@IEEEtitlepagestyle%
 \def\@oddfoot{\strut\hfill#1\hfill\strut}%
 \def\@evenfoot{\strut\hfill#1\hfill\strut}%
 }%
 \ps@headings%
}
\makeatother

\usepackage{soul}
\usepackage{multirow}
\usepackage{soul,color}
\usepackage{graphicx}
\usepackage{dsfont}
\usepackage{subcaption}

\usepackage[utf8]{inputenc}
\usepackage{mathtools}
\usepackage[mathscr]{euscript}
\usepackage{amsmath}
\usepackage{amssymb}
\usepackage{amsfonts}
\usepackage{verbatim}
\usepackage[T1]{fontenc}
\usepackage[utf8]{inputenc}

\usepackage{latexsym}
\usepackage{enumerate}
\usepackage{indentfirst}
\usepackage{calligra}
\usepackage{ulem}
\usepackage{multirow}

\usepackage{array}
\usepackage{float}
\usepackage{bbm}

\usepackage{eurosym}

\usepackage{hyperref}

\usepackage{caption}
\usepackage{tikz}
\usepackage{pgfplots}

\pgfplotsset{compat=1.8}
\usepgfplotslibrary{statistics}
\pgfmathdeclarefunction{fpumod}{2}{%
 \pgfmathfloatdivide{#1}{#2}%
 \pgfmathfloatint{\pgfmathresult}%
 \pgfmathfloatmultiply{\pgfmathresult}{#2}%
 \pgfmathfloatsubtract{#1}{\pgfmathresult}%
 \pgfmathfloatifapproxequalrel{\pgfmathresult}{#2}{\def\pgfmathresult{5}}{}%
 }

\usepackage{tabularx}
\usepackage{flushend}
\usepackage{textcomp}
\usepackage{xcolor}
\usepackage{import}

\usetikzlibrary{trees,decorations,shadows}
\tikzset{level 1/.style={sibling angle=45,level distance=4mm}}
\usetikzlibrary{arrows.meta}
\usepackage{forest}
\usetikzlibrary{external}

\let\oldtikzexternalgetnextfilename\tikzexternalgetnextfilename \renewcommand{\tikzexternalgetnextfilename}[1]{\oldtikzexternalgetnextfilename{#1}\expandafter\tikzsetnextfilename\expandafter{#1}}

\usepackage{pgfplotstable}
\usepackage[outline]{contour}
\contourlength{1.2pt}

\pgfplotsset{compat=1.13} 

\usetikzlibrary{spy}
\usetikzlibrary{calc}
\usetikzlibrary{fadings}
\usetikzlibrary{patterns}
\usetikzlibrary{shadows}
\usetikzlibrary{mindmap}
\usetikzlibrary{backgrounds}
\usetikzlibrary{shapes.symbols}
\usetikzlibrary{shapes.multipart}
\usetikzlibrary{shapes.geometric}
\usetikzlibrary{automata,positioning}
\usetikzlibrary{decorations.fractals} 
\usetikzlibrary{decorations.markings}
\usetikzlibrary{decorations.pathreplacing}
\usetikzlibrary{decorations.pathmorphing}

\usepackage{bm}

\usepackage{nomencl}
\makenomenclature
\tikzset{edge from parent/.style={segment angle=10,draw}}

\tikzset{
 my rounded corners/.append style={rounded corners=2pt},
}

\def\BibTeX{{\rm B\kern-.05em{\sc i\kern-.025em b}\kern-.08em
 T\kern-.1667em\lower.7ex\hbox{E}\kern-.125emX}}

\renewcommand{\nomgroup}[1]{%
 \ifthenelse{\equal{#1}{O}}{\item[\textit{Operators}]}{%
 \ifthenelse{\equal{#1}{I}}{\item[\textit{Indices}]}{%
 \ifthenelse{\equal{#1}{A}}{\item[\textit{Acronyms}]}{%
 `\ifthenelse{\equal{#1}{V}}{\item[\textit{Variables and parameters}]}{}}}}}
\usepackage{scalerel}
\usetikzlibrary{svg.path}
\definecolor{orcidlogocol}{HTML}{A6CE39}
\tikzset{
 orcidlogo/.pic={
 \fill[orcidlogocol] svg{M256,128c0,70.7-57.3,128-128,128C57.3,256,0,198.7,0,128C0,57.3,57.3,0,128,0C198.7,0,256,57.3,256,128z};
 \fill[white] svg{M86.3,186.2H70.9V79.1h15.4v48.4V186.2z}
 svg{M108.9,79.1h41.6c39.6,0,57,28.3,57,53.6c0,27.5-21.5,53.6-56.8,53.6h-41.8V79.1z M124.3,172.4h24.5c34.9,0,42.9-26.5,42.9-39.7c0-21.5-13.7-39.7-43.7-39.7h-23.7V172.4z}
 svg{M88.7,56.8c0,5.5-4.5,10.1-10.1,10.1c-5.6,0-10.1-4.6-10.1-10.1c0-5.6,4.5-10.1,10.1-10.1C84.2,46.7,88.7,51.3,88.7,56.8z};
 }
}

\newcommand\orcidicon[1]{\href{https://orcid.org/#1}{\mbox{\scalerel*{ \begin{tikzpicture}[yscale=-1,transform shape]
 \pic{orcidlogo};
 \end{tikzpicture}
 }{|}}}}

\begin{document}
%
\title{{Stochastic flexibility needs assessment: learnings from H2020 EUniversal's German demonstration}}

\author{Md~Umar~Hashmi*\orcidicon{0000-0002-0193-6703},
Simon~Nagels\orcidicon{0009-0008-6487-6526},
and~Dirk~Van~Hertem~\orcidicon{0000-0001-5461-8891}
\thanks{Corresponding author email: mdumar.hashmi@kuleuven.be}
\thanks{Md Umar Hashmi, Simon Nagels and Dirk Van Hertem are with KU Leuven, division Electa \& EnergyVille, Genk, Belgium}
\thanks{This work is supported by 
the \href{https://euniversal.eu/}{H2020 EUniversal project}, grant agreement ID: 864334 
and 
the Flemish Government and Flanders Innovation \& Entrepreneurship (VLAIO) through the IMPROcap project (HBC.2022.0733).
}} 

\maketitle

\begin{abstract}
Operational flexibility needs assessment (FNA) is crucial for system operators to plan/procure flexible resources in order to avoid probable network issues. We implemented an FNA tool in the framework of the H2020 EUniversal project for the German demonstration. In this work, we summarize our learnings from the demo implementation to cope with the limited availability of measurement data. Using a reduced network model and key performance indicators, we  evaluate the digital-twin results with real-world implementations. The paper aims to motivate future research directions by duly considering real-world limitations in their modelling and developing innovative tailor-made solutions for an improved decision support framework for system operators.
\end{abstract}


 \pagebreak

\tableofcontents

\pagebreak

\section{Introduction}

The evolution of distribution networks (DN) with greater amounts of load and distributed generation uncertainty makes it challenging for distribution system operators (DSOs) to operate the grid reliably. This energy transition demands developing decision support tools for the DSOs to use flexible resources while ensuring probable network congestion incidents and deteriorating power quality (power factor, voltage unbalance, etc.) events are minimized. The traditional fit-and-forget approach for the distribution networks needs a major revamping \cite{beckstedde2023fit}.
\vspace{-10pt}

\begin{figure}[!htbp]
	\center
	\includegraphics[width=6.2in]{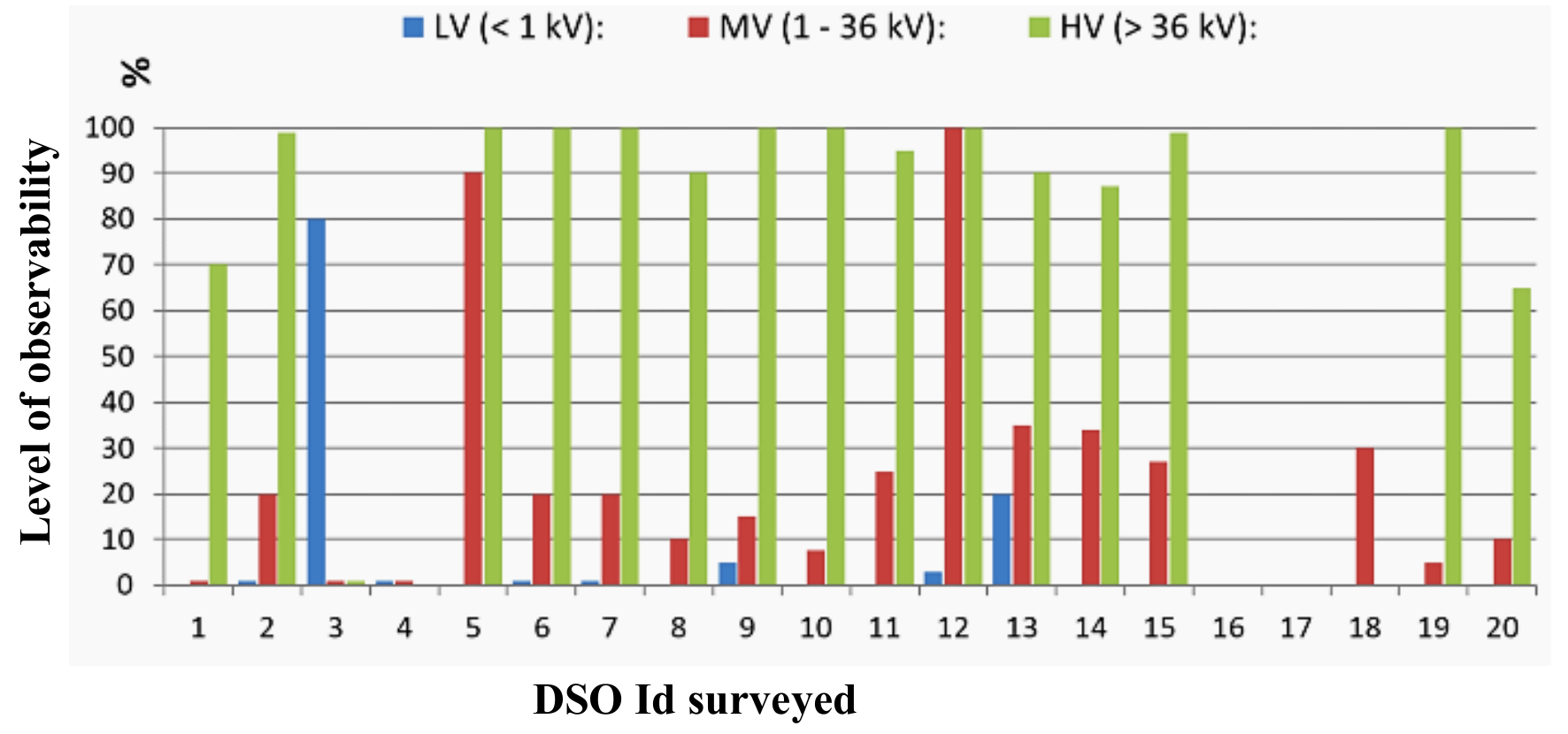}
	\vspace{-6pt}
	\caption{\small{Observability reported by DSOs \cite{website2}.}}
	\label{fig:edso}
\end{figure}

The primary goal of the EUniversal project is to address the current limitations in the utilization of flexibility by DSOs for managing congestion and the grid. In line with the European approach and the imperative for standardization and harmonization, one of EUniversal's key objectives is to establish and seamlessly integrate the Universal Market Enabling Interface (UMEI). 
The UMEI is evaluated at three distinct locations: Portugal, Germany, and Poland, where its effectiveness in procuring market-based flexibility is assessed across various use cases. The objective of this paper is to summarize the learnings from the implementation of our tool for flexibility needs quantification in the context of the German demonstration. 
These learnings aim to bridge the gap between digital-twin-based evaluation projected onto real-world implementation.
For the demo LV DNs, the observability is quite low, as also observed in the E-DSO report \cite{website2}.

\begin{figure}[!htbp]
	\center
	\includegraphics[width=6.1in]{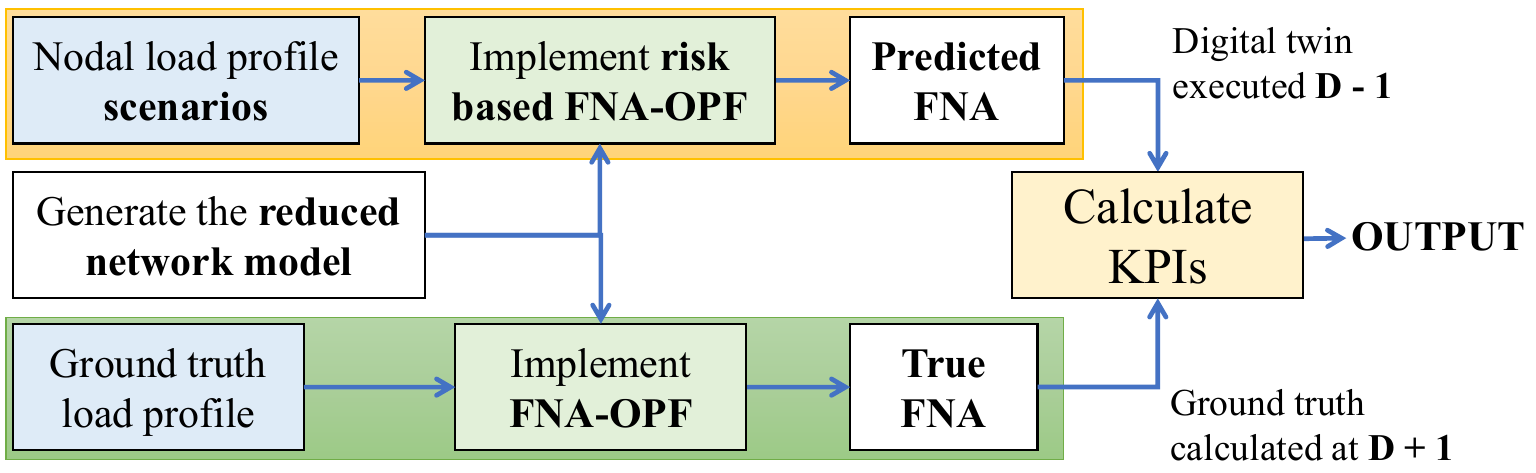}
	\vspace{-5pt}
	\caption{\small{Outline of FNA tool for demo evaluation and KPI calculation}}
	\label{fig:demooutline}
\end{figure}

The description of the LV Flexibility Needs Assessment (FNA) tool used in the German demo evaluation under the EUniversal project is introduced in \cite{hashmi2023robust}, and detailed also in the prior deliverables \cite{bockemuhl2022deliverable} and \cite{kratsch2023deliverable}. 
The network layout data extraction used in this demo is detailed in \cite{hashmi2023consensus}.
For the demo implementation of the FNA tool, the load profile scenarios are created using a prediction model based on historical data.
The demo networks have additional measurement devices installed, and home energy management devices are installed at prosumer locations opting to participate in the local flexibility market.
Thus, we employ a reduced DN model, see Fig. \ref{fig:demooutline}, for representing the demo DNs.
\vspace{-7pt}

\vspace{-4pt}

The key observations/contributions made in this work are
\begin{itemize}
\item DNs in Europe are often substantially oversized thus we redefine DN congestion to enable future readiness,
\item The limited observability in German DNs, motivates us to use a reduced network model. We propose a framework to reduce the network model based on the availability of the measurements. Further, we also provide a load aggregation mechanism that includes losses in the lines.
\item We perform stochastic optimization for considering parameter uncertainty, often leading to drastic over-prediction of flexibility needs of a distribution network. We observe that tuning this risk level is crucial to limit over-prediction while avoiding under-prediction of flexibility needs.
\item There is a need for tailor-made decision support tools for DSOs due to the unique problems faced by them. 
\end{itemize}

This paper is organized as follows.
Section \ref{section2}, outlines the online implementation of the flexibility needs quantification tool.
Section \ref{section3} details the network reduction and load scenario frameworks for the German demo networks.
Section \ref{section4} describes the key performance indicators (KPIs) used for assessing the performance of the demo outcome.
Section \ref{section5} details the demo results and calculated KPIs.
Section \ref{section6} summarizes the key takeaways of the paper.

\pagebreak

\section{Flexibility needs quantification}
\label{section2}
The Flexibility Needs Assessment (FNA) involves evaluating the level of flexibility required by the Distribution System Operator (DSO) to effectively plan and obtain flexibility from the market, minimizing the likelihood of Distribution Network Incidents (DNIs). 
DNIs encapsulate voltage limit violations and line overloads.
The FNA algorithm operates without assuming the specific locations of flexible resources. It assumes that flexibility is present at nodes with connected loads or generation sources. To model potential DNIs, uncertainties are incorporated using a Monte Carlo approach, simulating various scenarios based on nodal load and generation forecasts and historical forecast errors. For each scenario, an FNA-Optimal Power Flow (FNA-OPF) problem is solved. However, a robust FNA, considering the worst-case scenario, might lead to excessive procurement of flexibility. To mitigate this, a risk-based index, such as a chance constraint (CC), is introduced. Higher CC values indicate a greater risk the DSO may face in dealing with unresolved DNIs \cite{hashmi2023flexibility, hashmi2023robust, hashmi2023perspectives}.

\begin{figure}[!htbp]
    \center
    \includegraphics[width=5.1in]{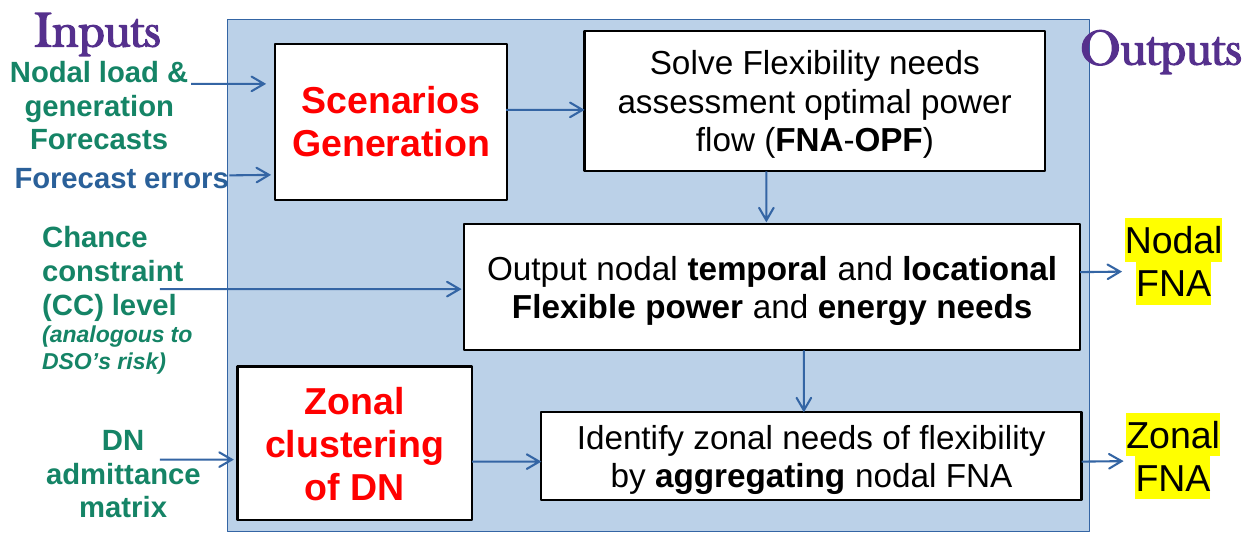}
    \vspace{-5pt}
    \caption{\small{The framework for flexibility quantification \cite{hashmi2023robust}.}}
    \label{fig:fnaframe}
\end{figure}

Fig. \ref{fig:fnaframe} shows the flexibility quantification framework utilized in the demo, originally proposed in \cite{hashmi2023robust}.

\vspace{-20pt}

\subsection{Demonstration implementation assumptions}
For the implementation of the FNA tool, we have made the following assumptions:\\
$\bullet$ \textit{Redefining the DNIs for demo DNs}: 
Since no DNIs are observed for the demo DNs, more restrictive limitations are imposed to verify the functioning of the FNA tool within the UMEI. The more stringent voltage limits and the loading limits of branches and transformers are included in Table \ref{tab:limits}. \\
$\bullet$ \textit{1-$\phi$ equivalent FNA}: From the consumer meta-data, we know that for the demo networks, more than 90\% of the consumers connected are single-phase consumers. Since the phase connectivity of the single-phase consumers is not known, thus, we assume a single-phase equivalent network layout for flexibility quantification. This simplification underestimates the true FNA.\\
$\bullet$ \textit{Network reduction}: The DN reduction proposed in this work simplifies the DN based on the availability of the measurements. 

\begin{table}[!htbp]
\centering
\caption{\small{Modified voltage and line loading limits for MLq0094 and MFn4420 DNs for tool analysis}}
\label{tab:limits}
\begin{tabular}{c|c|c}
Network           & MLq0094                    & MFn4420                    \\ \hline
Voltage limits    & 0.95-1.05 pu               & 0.95-1.05 pu               \\
Current limit     & 40\% of line rating        & 50\% of line rating        \\
Transformer limit & 50\% of transformer rating & 50\% of transformer rating
\end{tabular}
\end{table}

\vspace{-15pt}

\subsection{Online implementation}


The FNA tool consists of Python and Julia scripts. The set of Python scripts parses the networks from DigSilent to a readable format, creates reduced network files (see Section \ref{section3}) and generates scenarios for nodal load profiles from historical data. The set of Julia scripts is employed to calculate the FNA, i.e., the minimal need for flexibility in the DN while complying with line loading and voltage limit violations.  
Since no external data exchange is permissible, the FNA tool is implemented on the DSO server utilizing two \href{https://www.docker.com/}{docker containers}. 


\pagebreak

\section{Reduced network model for FNA}
\label{section3}

Limited observability in the demo networks, motivates us to form a reduced network model for flexibility quantification. The alternative option that is not considered in this work is load disaggregation, however, fragmenting load profiles is quite hard in this case due to less than 6.8\% of the nodes having measurement feedback. These measurements are not traditional smart meters, but rather three-phase measurement boxes, typically at feeder heads. In Germany, smart meter data is handled by meter data operators (MDO) and DSOs have limited access to this data due to privacy concerns. Thus, DSO installed additional measurements for the demo implementation, measuring line flows rather than consumer load, contrary to the smart meters. In this section, we detail the network reduction framework, and load scenarios and briefly describe the demo networks.

\subsection{Network reduction}
\begin{figure}[!htbp]
	\center
	\includegraphics[width=5.5in]{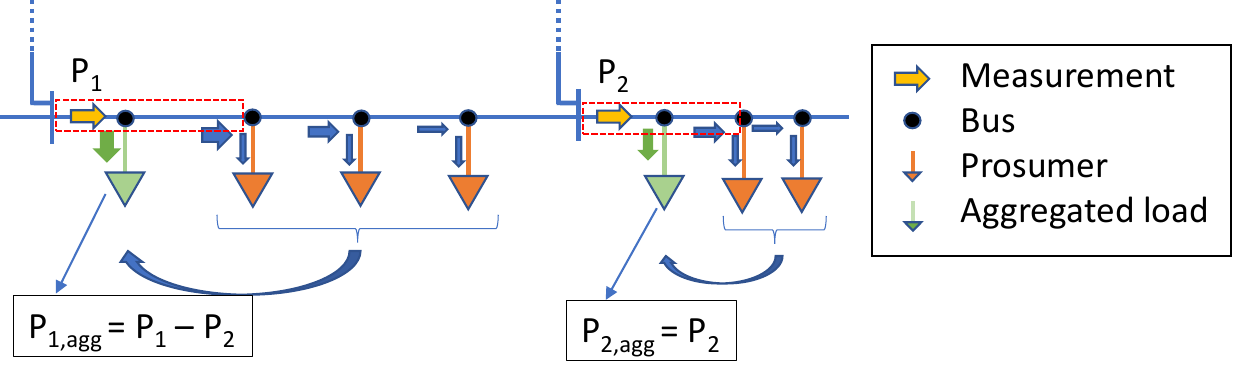}
	\vspace{-5pt}
	\caption{\small{Branch power flow conversion to load aggregation.}}
	\label{fig:b2}
\end{figure}

Due to limited measurement feedback, the original network is reduced in number of nodes and branches. This reduced network model uses aggregated load measurements for performing power flow studies.
Fig. \ref{fig:b2} shows the mechanism for network reduction.
The aggregated loads were placed at the first bus downstream of the measurement since in practice the power flow is the same between the measurement location and the first downstream prosumer, i.e., in the red box in Fig. \ref{fig:b2}. By placing the aggregated load downstream, the voltage and loading limits of the corresponding branch and bus are included in the flexibility calculation. This placement of the aggregated load is crucial since almost all available power measurements are located at the substation, therefore positioning the aggregated loads at the substation busbar itself would not include the loading limits of the feeders in the FNA calculation.

 \vspace{-8pt}

In Fig. \ref{fig:b2}, it is shown that the load between the measurement points 1 and 2, i.e., $P_{1,\text{agg}} = P_1 -P_2$, is aggregated at an upstream bus close by the measurement location. Since no downstream measurements exist for measurement point 2, the power flow measured at $P_2$ is aggregated at its first downstream bus.
The use of aggregated loads allows firstly to remove branches which are not loaded anymore, e.g., the branches downstream of $P_{2,\text{agg}}$ in Fig. \ref{fig:b2}. Secondly, branches can be merged and intermediate buses removed, e.g. applicable for the branches and nodes between $P_{1,\text{agg}}$ and the bus at $P_{2}$ in Fig. \ref{fig:b2}.  This reduction step retains the same electrical behavior (r, x) of the network, reduces the number of branches and nodes, and reduces the calculation time substantially. The rating of the merged branch is taken equal to the largest rating of all merged branches.


\subsection{Aggregated load profile scenarios}

Aggregated load profile scenarios were created using a persistence model and historic power flow measurements. 
The persistence model creates load profile scenarios by using the D-2 load 
as the mean expected load and a standard deviation equal to 30\% of the D-2 load. The historic aggregated loads are deduced from the historical measurements of the power flow at measurement locations. 
This method yields the 200 load profile scenarios used in the FNA tool.

\subsection{Demo networks}
\begin{figure}[!htbp]
	\center
	\includegraphics[width=5.5in]{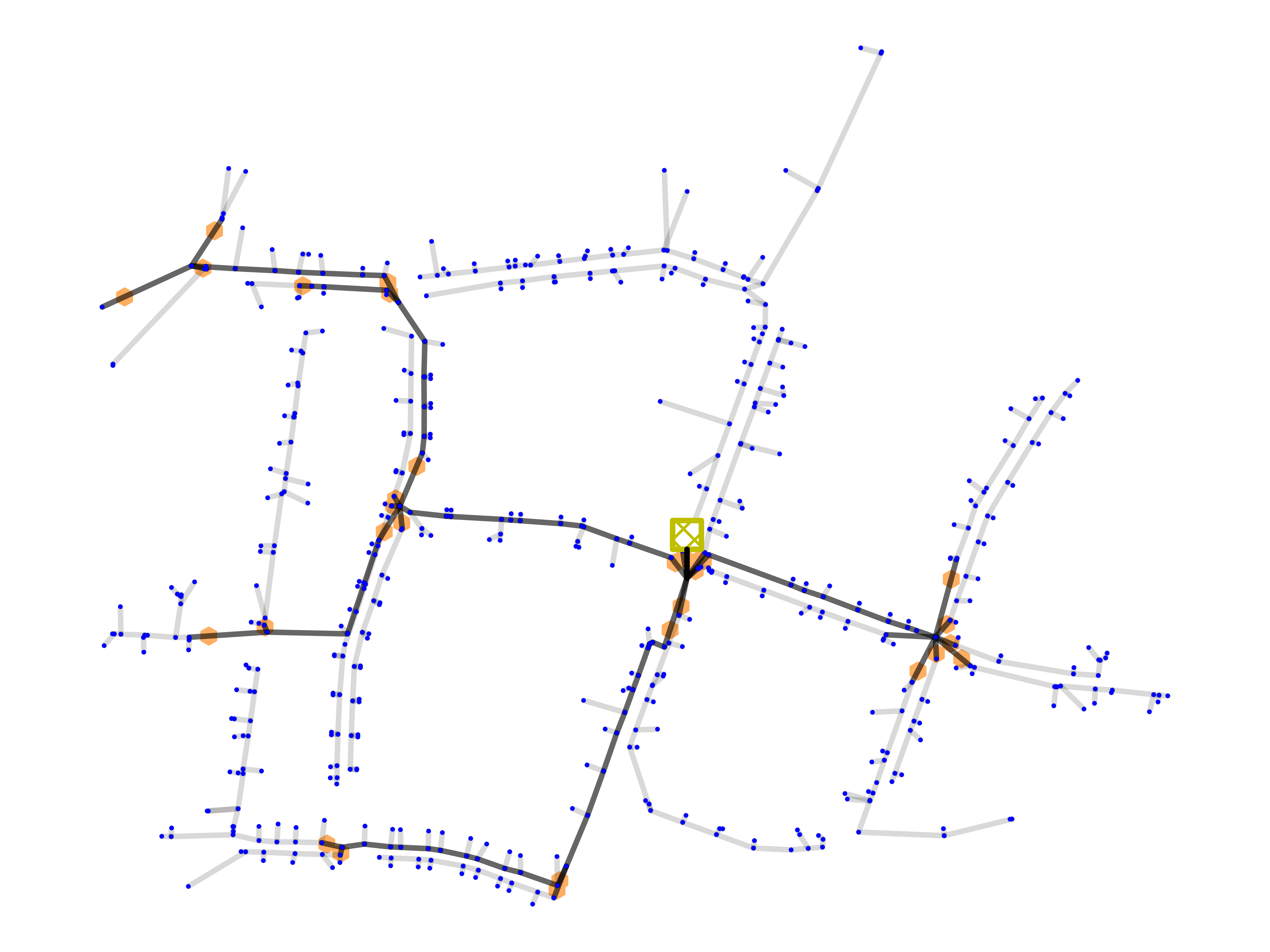}
	\vspace{-5pt}
	\caption{\small{EUniversal's demo network 1: MLq0094}}
	\label{fig:mlq0094}
\end{figure}

\begin{figure}[!htbp]
	\center
	\includegraphics[width=5.5in]{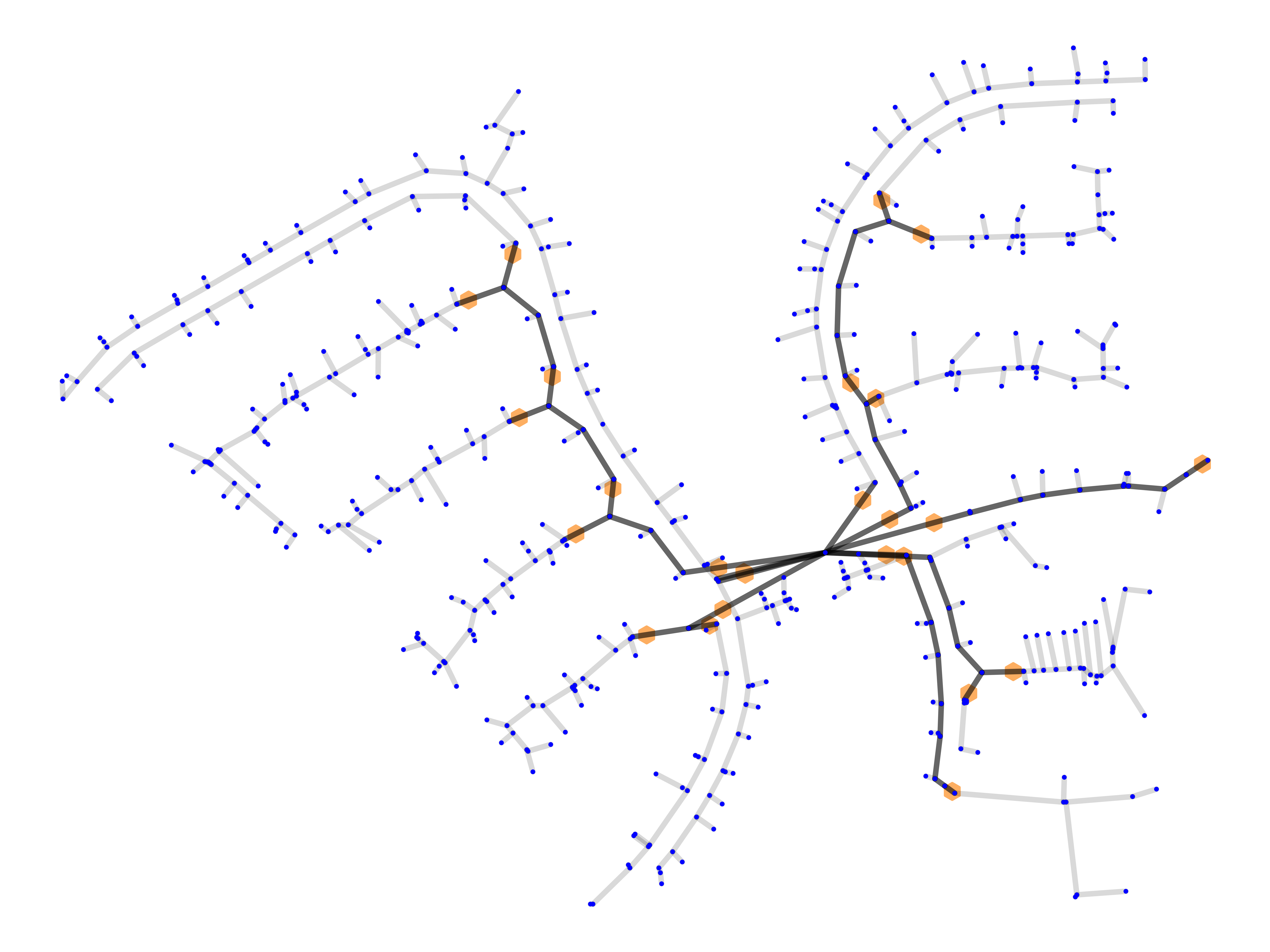}
	\vspace{-5pt}
	\caption{\small{EUniversal's demo network 2: MFn4420.}}
	\label{fig:mfn4420}
\end{figure}

Based on the power flow measurement locations in the DN, the reduced network models in Fig. \ref{fig:mfn4420} and \ref{fig:mlq0094} were derived. 
The blue topology shows the original DN and the black line shows the reduced DN with measurement locations shown with orange dots.
This reduction allows the employment of aggregated loads at the measurement locations and significantly reduces the tool computation time due to the reduced network size, see Tab. \ref{tab:my-table}. 

\begin{table}[!htbp]
\centering
\caption{Network size: original and reduced network}
\vspace{-8pt}
\label{tab:my-table}
\begin{tabular}{
>{\columncolor[HTML]{FFFFFF}}l 
>{\columncolor[HTML]{FFFFFF}}l 
>{\columncolor[HTML]{FFFFFF}}l 
>{\columncolor[HTML]{FFFFFF}}l 
>{\columncolor[HTML]{FFFFFF}}l }
              & \multicolumn{2}{l}{\cellcolor[HTML]{FFFFFF}MFn4420} & \multicolumn{2}{l}{\cellcolor[HTML]{FFFFFF}MLq0094} \\
              & Original                  & Reduced                 & Original                  & Reduced                 \\ \hline
Number of branches & 560                       & 32                      & 603                       & 40                      \\
Number of nodes      & 561                       & 38                      & 603                       & 40                      \\
Number of loads    & 222                       & 25                      & 331                       & 30                     
\end{tabular}
\end{table}

Current and power flow measurements, with a resolution of 15 minutes, are available for both the demo networks referred to as MFn4420 and MLq0094 networks at the substation and switch boxes in the network. Their location in the distribution network is presented in figures \ref{fig:mfn4420} and \ref{fig:mlq0094}. Although these measurements aid in determining the load in both networks, the observability of the network is limited, as only a limited number of branches are observed, thereby the measurements only capture the aggregated consumption and production of downstream prosumers. Note, that the measured quantities are not uniform across measurement locations. All available measurements include current, active power and reactive power for each phase except for the measurements at the substation of the MLq0094 network. These measurements only include the absolute value of the current measured per phase. To overcome the lack of power flow measurements in the MLq0094 network, the power flow is approximated by multiplying the current measurements with the nominal voltage until power flow measurements become available. This approximation firstly overestimates the power flow, since the measured current also includes a reactive power component. Secondly, only positive power flows are obtained (loads) since only the size of the current is measured and not its direction. Therefore, negative power flows caused by a net injection are considered as loads.

\subsection{FNA settings}
The FNA tool uses the reduced network model and load scenarios to calculate the need for flexibility in the network to avoid probable line loading and voltage limit violations. Since procuring flexibility for the most extreme scenarios would lead to a large over-procurement of flexibility, the FNA tool calculates the need for flexibility to avoid loading and voltage violations in 75\% of all load scenarios, i.e., a CC or risk Level of 0.25. This parameter was set to this value to avoid underestimations and large overestimation of the need for flexibility. 

The functioning of the FNA tool is showcased by discussing its results on the demo days (13th of September 2023). Since load and voltage limits are not violated in practice in the MFn4420 and MLq0094 networks, more stringent operational limits are imposed to validate the FNA tool. 
The redefined DNI settings used for the evaluation are detailed in Tab. \ref{tab:limits}.

The validation of the FNA tool during the demo days is performed by comparing the predicted need for flexibility with the actual need for flexibility, by comparing the load with the loading limits in the network, and by analyzing the voltage throughout the network. Both the temporal and locational aspects of the flexibility, the load and loading limits are analyzed. The exact need for flexibility during a demo day is deduced after the demo day by employing the power measurements of those days to recreate the actual power flow in the test network.   

\vspace{-13pt}

\vspace{-20pt}


\pagebreak

\section{KPI calculation framework}
\label{section4}
\begin{figure}[!htbp]
	\center
	\includegraphics[width=5.8in]{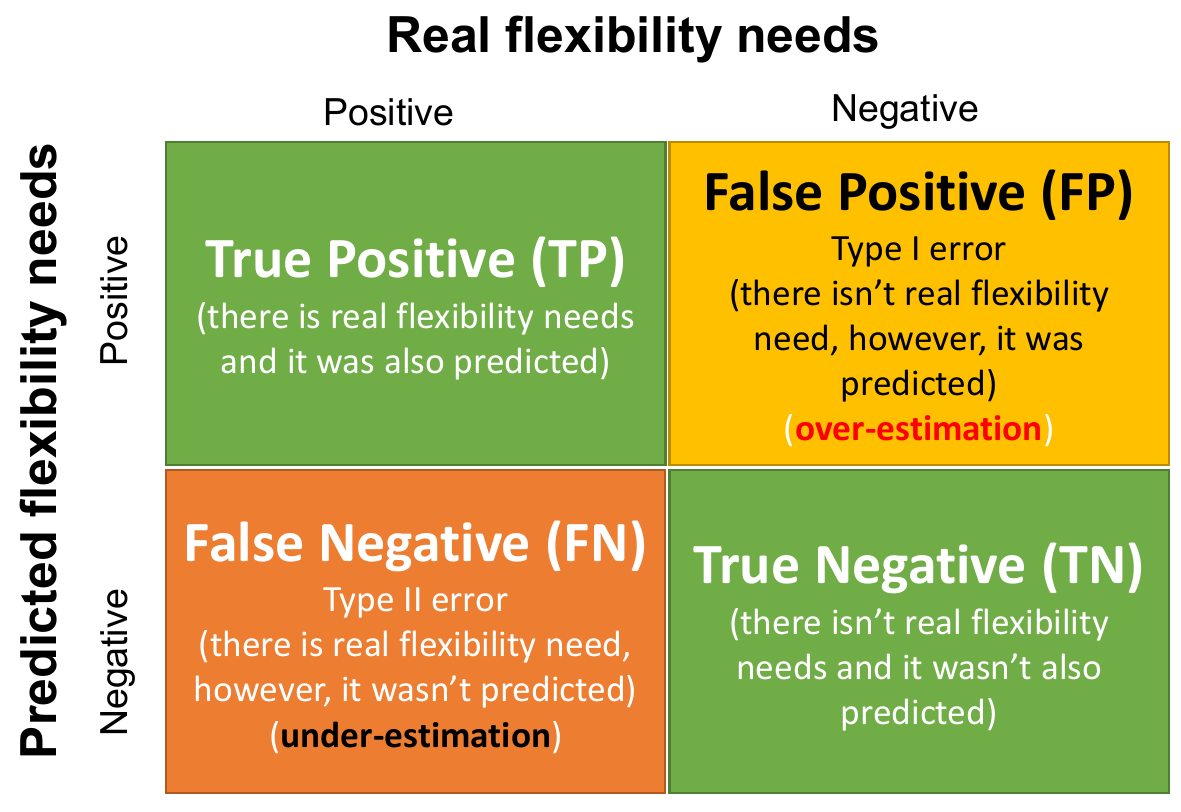}
	\vspace{-5pt}
	\caption{\small{Confusion matrix in the context of FNA.}}
	\label{fig:confusion}
\end{figure}

Fig. \ref{fig:confusion} shows the comparison of actual and predicted FNA. 
The accuracy of the FNA tool predicting future flexibility needs is assessed by employing three Key Performance Indicators (KPIs). These KPIs quantify the over and underestimation of the flexibility needs for different levels of risk, and employ the locational (bus; b = 1...N) and temporal (time; t=1,..,24) flexibility aspects of the predicted (denoted as $\text{Flex}_{\text{Pred}}^{(t,b)}$) and the actual (denoted as $\text{Flex}_{\text{Act}}^{(t,b)}$) need for flexibility.
Ideally, we would like to select a risk level that minimizes false negative (type II error) instances while also reducing false positive (type I error) instances.
From Fig. \ref{fig:confusion}, observe that a high type I error would lead to over-estimation of FNA, while a high type II error would lead to underestimation of FNA.
It is expected that stochastic FNA would lead to overestimation for ensuring that the type II error is minimized, as missing a true DNI is not desired. 
\vspace{-10pt}

$\bullet$~The S1 KPI, a true-positive KPI, denotes how often the actual need for flexibility is met by the predicted need for flexibility, in both temporal and locational terms. \vspace{-5pt}    
    \begin{equation} \label{KPI_S1}
        S1 [\%] = \frac{\sum_{t,b}^{}\mathbbm{1}(|\text{Flex}_{\text{Pred}}^{(t,b)}| \geq |\text{Flex}_{\text{Act}}^{(t,b)}|)}{\sum_{t,b}^{} \mathbbm{1}(|\text{Flex}_{\text{Act}}^{(t,b)}| > 0)} \times 100
    \end{equation}

$\bullet$~The S2 KPI, a true-positive KPI, denotes the overestimation of the need for flexibility in kW for all the actual locational and temporal flexibility needs.     
    \begin{equation}
        S2 [kW] = \sum_{t,b}^{}\mathbbm{1}(\text{Flex}_{\text{Act}}^{(t,b)} > 0)
|\text{Flex}_{\text{Pred}}^{(t,b)}| - |\text{Flex}_{\text{Act}}^{(t,b)}|.
    \end{equation}

$\bullet$~The S3 KPI, encompassing true positive and false positive incidents, denotes the total overestimation of the need for flexibility, thereby also including the overestimation if no flexibility was required in locational and temporal terms. This KPI is expressed in relative terms [\%] if an actual need for flexibility was observed. 
    \begin{equation}
    S3 = 
    \begin{dcases}
    \frac{\sum_{t,b}^{}|\text{Flex}_{\text{Pred}}^{(t,b)}| - |\text{Flex}_{\text{Act}}^{(t,b)}|}{\sum_{t,b}^{} |\text{Flex}_{\text{Act}}^{(t,b)}|},~\text{if} \sum_{t,b}^{} |\text{Flex}_{\text{Act}}^{(t,b)}| > 0 
    \\
    \sum_{t,b}^{}|\text{Flex}_{\text{Pred}}^{(t,b)}|, ~~\text{if} ~~\sum_{t,b}^{} |\text{Flex}_{\text{Act}}^{(t,b)}| =  0
    \\
    \end{dcases}
    \end{equation}    


\pagebreak

\section{Demonstration results}
\label{section5}
For the demo networks shown in Fig. \ref{fig:mlq0094} and \ref{fig:mfn4420}, stochastic FNA-OPF is implemented. The demo is implemented for the 13th of September 2023. The predicted FNA is identified on the 12th of September, with measurements accumulated for the 11th of September.
Next, we present the results for the demo networks and KPIs calculated for the demo day.
\vspace{-15pt}

\subsection{Demo network 1 (MLq0094)}
\begin{figure}[!htbp]
	\center
	\includegraphics[width=5in]{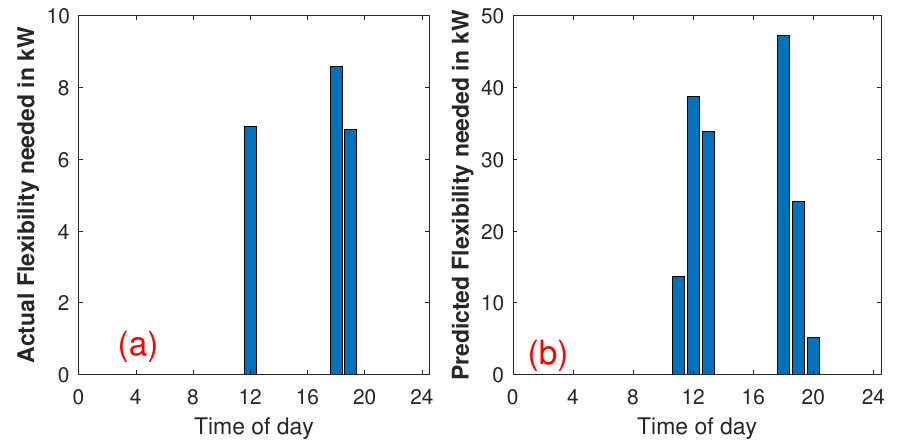}
	\vspace{-5pt}
	\caption{\small{Actual and Predicted Flex Needs [kW] on 13 September in MLq0094 network: temporal variation}}
	\label{fig:FNA_mlq_13_time}
\end{figure}

\begin{figure}[!htbp]
	\center
	\includegraphics[width=5in]{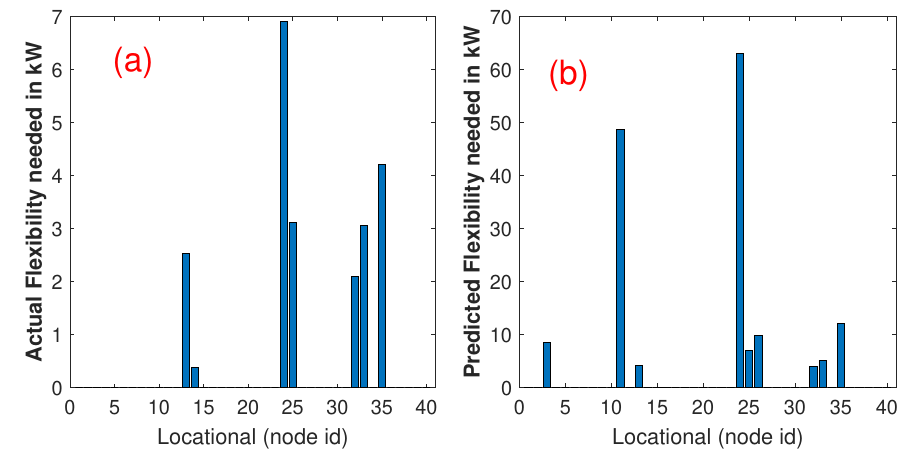}
	\vspace{-5pt}
	\caption{\small{Actual and Predicted Flex Needs [kW] on 13 September in MLq0094 network: locational variation}}
	\label{fig:FNA_mlq_13_locational}
\end{figure}

For MLq0094, Figures \ref{fig:FNA_mlq_13_time} and \ref{fig:FNA_mlq_13_locational} show the quantified flexibility comparison for the actual and predicted case in terms of temporal and locational variations.
Note that the predicted FNA is substantially higher than the actual FNA needed by the DSO in real time. One of the primary reasons for the disparity between predicted and actual FNA pertains to a network bottleneck caused by the tie line connecting the branching feeders from the substation. We recommend the DSO upgrade this line with a parallel conductor as in this case, FNA requirements are too high.

\vspace{-15pt}

\begin{figure}[!htbp]
	\center
	\includegraphics[width=6.3in]{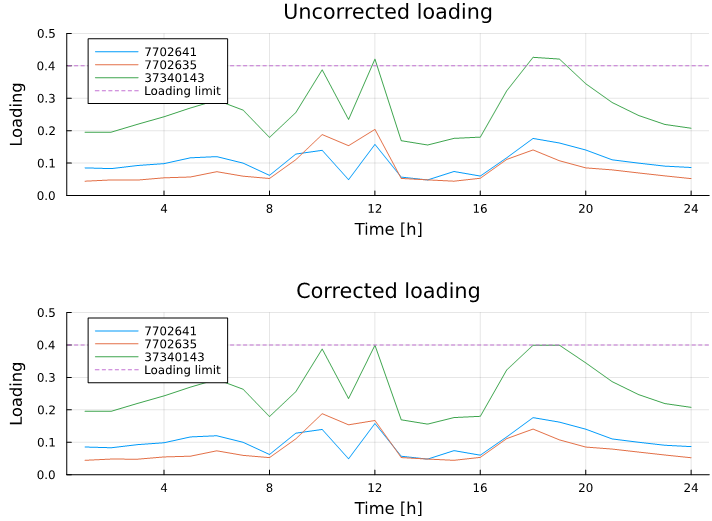}
	\vspace{-5pt}
	\caption{\small{FNA tool impact: (a) uncorrected network line loading versus (b) corrected line loading.}}
	\label{fig:test}
\end{figure}

Fig. \ref{fig:test} shows the line loading observed prior to flexibility quantification. In the second plot, observe the FNA tool activates some flexibility to avoid the DNI previously observed.




\subsection{Demo network 2 (MFn4420)}

\begin{figure}[!htbp]
	\center
	\includegraphics[width=6.5in]{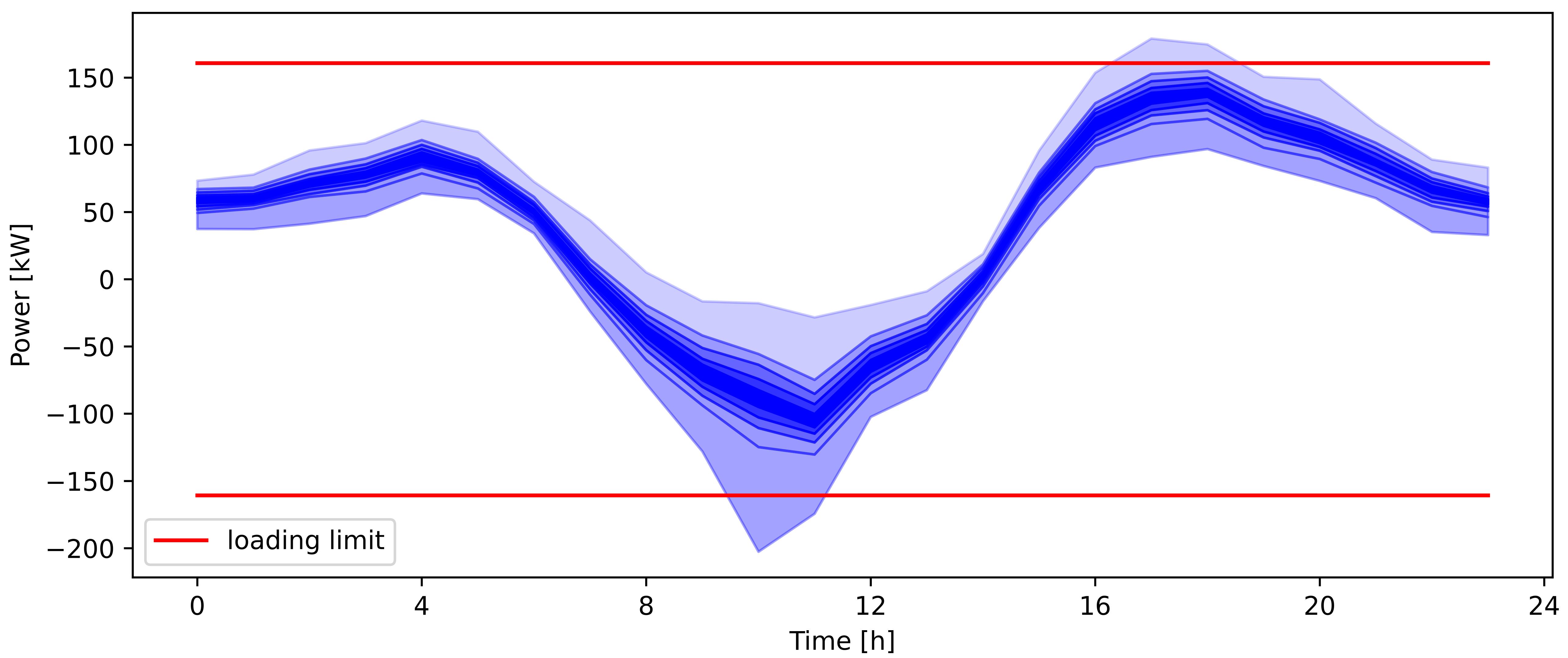}
	\vspace{-5pt}
	\caption{\small{Aggregated load profile scenarios used for Flex Needs Assessment on 13 September in MFn4420 network}}
	\label{fig:LoadMFn}
\end{figure}

For the second demo network, we observed that no flexibility is predicted and actually needed for a CC level of 0.25.
Fig. \ref{fig:LoadMFn} shows the loading scenario distribution, with the shade of blue getting darker for every 10\% increase in quartile close to the mean.
Note that this DN accommodates a substantial portion of PV generation, leading to reverse power flow during the day. The loading scenarios exceed the line limits momentarily for extreme worst-case scenarios, this is reflected in Tab. \ref{tab:kpi_mfn}.



Note that the reduced model of the demo DNs with most of the measurements placed close to the substation is unable to capture the voltage fluctuations at the end of the feeder due to low voltage drop. The placement of additional measurements close to the end of the feeders would substantially improve the capturing of voltage limit violations.

\subsection{KPI calculations for demo networks}

The KPI calculation for the demo day (13th of Sept.)
are listed in tables \ref{tab:kpi_mlq} and \ref{tab:kpi_mfn}.
Note that for the case of the demo DN 1, the estimated FNA is substantially higher compared to the actual FNA. This is due to the low current carrying capacity of the tie line connecting the substation and all the branching-out feeders, thus leading to very high levels of overestimation. In this case, the DSO should consider upgrading this line rather than planning for flexibility.
For the other demo DN, the actual FNA is zero while due to consideration of uncertainty, the DSOs are advised to plan a small level of flexibility in the form of generation curtailment during the day.
From tables \ref{tab:kpi_mlq} and \ref{tab:kpi_mfn}, note that the proposed tool is able to capture more than 93\% of true-positive FNA instances accurately for a risk level of 0.3 or lower.



\begin{table}[!htbp]
\caption{\small{KPI calculation for MLq0094 DN for different risk levels}}
\vspace{-7pt}
\label{tab:kpi_mlq}
\resizebox{\columnwidth}{!}{\begin{tabular}{lccccccccccc}
                                 & \multicolumn{11}{c}{Risk level}                                                                                                                                                                                     \\ \cline{2-12} 
\multicolumn{1}{l|}{}            & \textbf{0} & \textbf{0.05} & \textbf{0.1} & \textbf{0.15} & \textbf{0.2} & \cellcolor[HTML]{FFFFFF}\textbf{0.25} & \cellcolor[HTML]{C0C0C0}\textbf{0.3} & \textbf{0.4} & \textbf{0.5} & \textbf{0.6} & \textbf{0.7} \\
\multicolumn{1}{l|}{S1 (\%)}          & 100        & 93            & 93           & 93            & 93           & \cellcolor[HTML]{FFFFFF}93            & \cellcolor[HTML]{C0C0C0}93           & 71           & 50           & 21           & 7            \\
\multicolumn{1}{l|}{S2 {[}kW{]}} & 597        & 365           & 306          & 249           & 209          & \cellcolor[HTML]{FFFFFF}181           & \cellcolor[HTML]{C0C0C0}155          & 103          & 67           & 26           & 2            \\
\multicolumn{1}{l|}{S3 (\%)}     & 3818       & 1823          & 1301         & 1003          & 795          & \cellcolor[HTML]{FFFFFF}630           & \cellcolor[HTML]{C0C0C0}489          & 256          & 136          & 45           & -39         
\end{tabular}}
\end{table}

\begin{table}[!htbp]
\caption{\small{KPI calculation for MFn4420 DN for different risk levels}}
\vspace{-7pt}
\label{tab:kpi_mfn}
\resizebox{\columnwidth}{!}{\begin{tabular}{lccccccccccc}
                                 & \multicolumn{11}{c}{Risk level}                                                                                                                                                                                     \\
                                 & \textbf{0} & \textbf{0.05} & \textbf{0.1} & \textbf{0.15} & \textbf{0.2} & \cellcolor[HTML]{C0C0C0}\textbf{0.25} & \cellcolor[HTML]{FFFFFF}\textbf{0.3} & \textbf{0.4} & \textbf{0.5} & \textbf{0.6} & \textbf{0.7} \\ \cline{2-12} 
\multicolumn{1}{l|}{S1 (\%)}          & 100        & 100           & 100          & 100           & 100          & \cellcolor[HTML]{C0C0C0}100           & \cellcolor[HTML]{FFFFFF}100          & 100          & 100          & 100          & 100          \\
\multicolumn{1}{l|}{S2 {[}kW{]}} & /          & /             & /            & /             & /            & \cellcolor[HTML]{C0C0C0}/             & \cellcolor[HTML]{FFFFFF}/            & /            & /            & /            & /            \\
\multicolumn{1}{l|}{S3 {[}kW{]}}     & 164        & 48            & 22           & 10            & 3            & \cellcolor[HTML]{C0C0C0}0             & \cellcolor[HTML]{FFFFFF}0            & 0            & 0            & 0            & 0           
\end{tabular}}
\end{table}

\vspace{-20pt}

\pagebreak

\section{Key takeaways}
\label{section6}

The key takeaways from the demo implementation of the flexibility needs assessment (FNA) tool in the EUniversal project are as follows:\\
$\bullet~$	\textit{Tool competency}: The FNA tool provides the temporal and locational flexibility needs for the distribution network. The system operator can utilize this information for flexibility planning in operational timescales via the local flexibility market.\\
$\bullet~$	\textit{Need for improved forecasting}: the FNA tool is sensitive to the quality of forecast of nodal load profiles. In this work, we utilize a persistence model for generating load profile scenarios using D-2 load profiles. We use 30\% variance levels to accommodate uncertainty. Improved forecast models would allow predicting more accurately the FNA, thereby yielding smaller overestimations while eliminating Type II errors.\\
$\bullet~$	\textit{Setting of the risk level}: We compare the forecasted FNA with the ground truth FNA of the DN. The ground truth FNA is calculated using the true measurements for the demo day under consideration. The risk levels assist system operators in considering future uncertainties while avoiding over-estimating the true flexibility needs of the DN. Using the demo networks for the selected demo days, we observe that the risk value needs to be adjusted between 10 to 30\% to maximize the accurate estimation of FNA while reducing the overestimation of FNA. \\
$\bullet~$	\textit{Flexibility needs assessment vs network upgrade}: For the demo network MLq0094, we observe very high levels of predicted flexibility. This is primarily due to the bottleneck caused by the line connecting the transformer with the feeder bus bar. For this case, the efficient step to take is to upgrade the line with a higher ampacity. For such instances, avoiding infrastructural upgrades would not be recommended. A more detailed assessment needs to be performed to quantify the economic value of the line upgrade.\\
$\bullet~$	\textit{Need for more measurements}: Increased levels of the availability of the network measurements will increase the performance of the FNA tool by allowing for a less reduced network and increasing the accuracy of the load profile scenarios. Increased observability would also disaggregate the loads, which otherwise lead to underestimation of DNIs due to aggregation.

\pagebreak

\bibliographystyle{IEEEtran}
\bibliography{reference}

\end{document}